\documentstyle[12pt]{article}
\def\beq{\begin{equation}}   \def\eeq{\end{equation}}
\begin{document}
\begin{titlepage}

\begin{flushright}
HUTP-99/A020\\
NUB 3197\\
EFI-99-12\\
June 30, 1999
\end{flushright}

\vspace{0.3cm}

\begin{center}
\baselineskip25pt

{\Large\bf Higgs as a Slepton}

\end{center}

\vspace{0.3cm}

\begin{center}
\baselineskip12pt

{\large Aaron K. Grant$^1$ and 
Zurab Kakushadze$^{1,2,3}$}

\vspace{0.3cm}
$^1$Jefferson Laboratory of Physics, Harvard University,
Cambridge,  MA 02138\\
\vspace{0.3cm}
$^2$Department of Physics, Northeastern University, Boston, MA 02115\\
\vspace{0.3cm}
$^3$Enrico Fermi Institute, University of Chicago, Chicago, IL 60637\\ 
\vspace{0.3cm}
{\it grant@gauss.harvard.edu, 
zurab@string.harvard.edu}

\vspace{2.5cm}

{\large\bf Abstract} 

\vspace*{.25cm}

\end{center}
In theories with TeV scale quantum gravity there is a logical possibility
where the electroweak Higgs can be a fourth generation slepton. Despite
maximal supersymmetry breaking such a scenario turns out to be tightly
constrained, and might be testable already at present collider
experiments. Also, a light Higgs is no longer a prediction of
supersymmetry.
\end{titlepage}

{}Recently it was suggested that the fundamental scale of quantum gravity
$M_{Pf}$ may be as low as a few TeV, which would provide an alternative way
of understanding the hierarchy problem \cite{add}. In this proposal the
observed weakness of gravity at large distances is due to the presence of
$n$ large new dimensions (of size $\sim R >> M_{Pf}^{-1}$) in which gravity
can propagate.  The relation between the observed Planck scale $M_P$ and
the fundamental Planck scale $M_{Pf}$ is given by
\begin{equation}
M_P^2 \sim  M_{Pf}^{n + 2}V_n~, \label{planckscale}
\end{equation}
where $V_n \sim R^n$ ($n$ must be 2 or larger) is the volume of the $n$ extra
spatial dimensions ({\em i.e.}, of the bulk). In this picture
all the Standard Model fields must reside inside a brane 
(or a set of branes) with $3$ extended spatial 
dimensions\footnote{In a somewhat different
context an attempt to lower the string scale down to a TeV (without lowering
the fundamental Planck scale) was considered in \cite{lyk} based on an
earlier observation in \cite{witten}.}. A general discussion of possible
embeddings of such a scenario within the string theory context was given in 
\cite{aadd,st,kt}.

{}In this framework the role of supersymmetry might be somewhat 
modified. {\em A priori} it is no longer needed as a solution to the 
hierarchy problem. However, it might still play an important role in, say, 
formulating the fundamental theory that incorporates quantum gravity above a
TeV. At present the only known candidate theory for such a unified picture is
superstring theory. Thus, we may expect string theory to take over at  
energies above a TeV with supersymmetry broken not much below the fundamental 
string scale $M_s$. If so, what is the content of the supersymmetric theory 
just below $M_s$ where stringy modes can be integrated out?

{}The purpose of this note is to point out that in theories with quantum
gravity (or strings) at the TeV scale, there is a logical possibility to
identify the electroweak Higgs with the scalar superpartner of one of the
lepton doublets (or a linear combination thereof).  However, we cannot
identify the $SU(2)$-breaking slepton with the superpartner of a known
lepton.  Rather, we propose the existence of a fourth generation, whose
slepton plays the role of the Higgs boson.  In this way, the Higgs and
flavor structures of the theory are more closely related than in the usual
supersymmetric standard model.  Other proposals have been made wherein a
sneutrino belonging to one of the first three generations has a small
$SU(2)$-breaking vacuum expectation value \cite{Liu}.  Our proposal differs
in that the {\it fourth} generation sneutrino is the {\it sole} source of
$SU(2)$ breaking.

{}Already in the early days of supersymmetric model building, the
intriguing similarity of the quantum numbers of the weak lepton doublets
($L^a$) and the electroweak Higgs $H$ made it tempting to speculate that
the Higgs might actually be the superpartner of a lepton (that is, a
slepton). Unfortunately, within the usual paradigm with the fundamental
Planck scale $\sim 10^{19}$ GeV this is impossible because of the well
known rules for writing supersymmetric Lagrangians. Thus, the couplings
\begin{equation}
  L^* Q U_c~,
\end{equation}
which in these scenarios are required to give masses to the up-type
quarks, would be suppressed by the factor $M_{susy}^2/M_P^2$, which can at
best be $\sim 10^{-32}$ or so.

{}However, the situation is dramatically different if the fundamental
Planck scale is around a TeV. Now the desired couplings of the slepton
Higgs to the fermions can be generated through the couplings in the
K\"ahler potential
\begin{equation}
\int d^4\theta g_{ab}^u{X^* L^* \over M_{Pf}^2}Q^aU_c^b ~,
\end{equation}
where $X$ is the ``spurion'' superfield which breaks supersymmetry through
the F-term $\langle F_X \rangle\sim \theta^2 M_{Pf}^2$. Then the above
terms generate effective Yukawa couplings of the form
$g_{ab}^u L_{\rm scalar}^* Q_{\rm fermion}^a U_{\rm c,fermion}^b$,
where $L$ is the lepton superfield whose scalar component is identified
with the Higgs, and $U_c$ is the charge-conjugated up-type quark. Note that
in this context no additional suppression appears.  The couplings that
generate masses of the down quarks and charged leptons can come (depending
on the details of a given model) from the superpotential (as usual)
\begin{equation}
\Delta W = g_{ab}^dL Q^aD^b + g_{ab}^eL L^ae_c^b~,\label{wdownm}
\end{equation}
and/or from the K\"ahler potential
\begin{equation}
\Delta K = {X^* \over M_{Pf}^2} (g_{ab}^dL Q^aD^b + g_{ab}^eL
L^a e_c^b)~.\label{kdownm}
\end{equation}
Small neutrino masses (if they are non-zero) are more subtle and
may involve intrinsically higher dimensional mechanisms \cite{addm} which
will not be discussed here. Here we should point out that the mass of the 
fermionic component of $L$ (that is, of the lepton superfield whose scalar
component is identified with the Higgs) {\em cannot} be generated from the 
above couplings. Indeed, couplings $LLe^a_c$ and $X^* LLe^a_c$ vanish due
to the antisymmetry of the weak $SU(2)$ contractions. However, there are
other terms in the K\"ahler potential that give rise to the desired Yukawa
couplings $L_{\rm scalar} L_{\rm fermion} e^a_{\rm c,fermion}$. 
For instance, consider the term
\begin{equation}\label{superD}
 {({\cal D}X)X^*\over M_{Pf}^4}L({\cal D}L)
 e^a_c~,
\end{equation}
where ${\cal D}$ denotes the covariant superderivative.
In fact, if there is some hierarchy between the supersymmetry breaking
scale and $M_{Pf}$, then this might be (partially) responsible for the
fermion mass hierarchy between the quark and lepton sectors.

{}{\em A priori} within this framework there might arise problems
associated with lepton flavor violation. However, this problem might be
solved by introducing appropriate flavor symmetries. We specialize to the
case where the electroweak Higgs is identified with a fourth generation
slepton $L^4$.  (It will become clear momentarily why we cannot identify
the Higgs boson with the superpartner of a known neutrino.) Then the couplings
(\ref{wdownm}) that generate charged lepton masses will also generate the
couplings of the form
\begin{equation}
g_{ab}^e L^4_{\rm fermion} L^a_{\rm scalar}e^b_{\rm c,fermion}~,
\label{couplings}
\end{equation}
which will generically induce various flavor violating transitions.  The
same will be true if masses are generated from (\ref{kdownm}).  Thus, we
have to impose flavor symmetries. For instance, we can attempt to impose
individual $e$-, $\mu$- and $\tau$-lepton number symmetries.  This would
guarantee that the individual lepton numbers are conserved. Alternatively,
we may attempt to solve the flavor violation problem along the lines of
\cite{bd,zk}\footnote{Fermion Yukawa hierarchy from flavor symmetry
breaking on distant branes \cite{AD} can also be combined with this
scenario.}.  Here we should point out that the baryon number violating
terms must be adequately suppressed to avoid too rapid proton
decay. Generically this requires imposing additional symmetries (see, {\em
e.g.}, \cite{zk1}). However, it appears that this can be done consistently
with generating acceptable fermion masses in the above framework \cite{dk}.

The spectrum of new particles in this class of models differs from that
found in the minimal supersymmetric standard model.  In particular, the
model has only a single Higgs boson and only three neutralinos.  The
charginos of the model are a linear combination of the charged gauginos and
the fourth generation lepton.  Their mass matrix has the form
\begin{equation}
\left(
\begin{array}{cc}
       M_2           & 0   \\
        \sqrt{2} m_W & m_4
\end{array}
\right)
\end{equation}
in the $(\widetilde{W} , L_4)$ basis, where $M_2$ is the $SU(2)$ gaugino
mass, and $m_4$ is the fourth generation lepton mass.  The neutralinos are
mixtures of the neutral gauginos and the fourth generation neutrino.  In
the $(\widetilde{B}, \widetilde{W}^3, \nu_4)$ basis, their mass matrix has
the form
\begin{equation}
\left(
\begin{array}{ccc}
       M_1           & 0                & -m_Z \sin\theta_W \\
       0             & M_2              &  m_Z \cos\theta_W \\
   -m_Z\sin\theta_W  & m_Z \cos\theta_W &  0                \\
\end{array}
\right).
\end{equation}
We can now explain why it is impossible to identify the Higgs boson with
the superpartner of a known neutrino.  Indeed, had we identified the
Higgs doublet with, say, the tau lepton doublet, then we would find that
tau neutrino typically acquires a see-saw type mass of order $m_Z^2
/M_{1,2}$.  This mass, of order 10 GeV, is far too large.  Hence we choose
the alternate route of introducing a fourth generation, and identifying
{\it its} sneutrino with the Higgs boson.

Limits from collider searches for new particles constrain the parameter
space of this class of models.  The charginos have masses that can exceed
LEP-II collider limits \cite{Expts}, provided that $M_2$ and $m_4$ are
sufficiently large.  However, the neutralinos cannot be made arbitrarily
heavy, and indeed this class of models always has at least one relatively
light neutralino.  As an example, take the values
\begin{equation}
M_1 = 45~{\rm GeV},~~~M_2 = 110~{\rm GeV},~~~m_4 = 185~{\rm GeV}.
\end{equation}
We then have $m_{\chi^{+}_{1,2}}=90,~226$ GeV, while the neutralino masses
are
\begin{equation}
m_{\chi^0_1} = 56~{\rm GeV},~~~m_{\chi^0_2} = 57~{\rm GeV},~~~m_{\chi^0_3} =
156~{\rm GeV}.
\end{equation}
The lightest neutralino is predominantly $U(1)$ gaugino, the next to
lightest is mostly neutrino, and the heaviest is mostly $SU(2)$ gaugino.
In $e^+e^-$ collisions at $\sqrt{s}=183$ GeV, the production cross sections
for the lightest two neutralinos are 0.002 pb and 0.31 pb respectively,
provided that the sfermions are heavy so that $Z$ exchange is the dominant
production mechanism.  The relative size of the cross sections is easy to
understand, since the lightest neutralino is mostly bino and therefore
couples very weakly to the $Z$.  In this particular case, neutralino
production will be dominated by $\chi^0_2$ pairs.  The $\chi^0_2$ may
either decay promptly to three-body final states, or may decay first to
$\chi^0_1 Z^*$ or $\chi^0_1 \tilde{\nu}_4^*$, leading to five (or
more)-body final states.  The decays of the $\chi^0_{1,2}$ are highly model
dependent, but we can delineate the main possibilities.

\begin{enumerate}
\item
R-parity is strictly preserved for the first three generations, and broken
only by the Yukawa couplings of the fourth generation lepton.  Here we have
two possibilities, depending on the mass of the gravitino $\tilde{G}$.  If
decays to the gravitino are allowed, we expect decays like $\chi^0_{1,2}
\rightarrow \bar{f} f \tilde{G}$, where $f$ can be a lepton or quark.  If
decays to the gravitino are kinematically forbidden, the lightest
neutralino is absolutely stable.  In this case, $\chi^0_2$ pair production
will result in events with rather soft jets and leptons coming from the
$\chi^0_2\rightarrow\chi^0_1 Z^*,~\chi^0_1 \tilde{\nu}_4^*$ decay. Also note
that the lightest neutralino in this case is the SUSY dark matter
candidate.

\item
If R-parity is broken for all four generations, then $\chi^0_{1,2}$ will
decay to the standard model particles.  The decays are mediated by
couplings of the form $L L e_c$ or $L Q D$, where the superfields can
belong to any of the four generations.  The cross section for $\chi^0_2$
pair production for the sample point given above is 0.31 pb.  This exceeds
the experimental limit of roughly 0.1 pb \cite{Expts} if the neutralino
decays primarily to leptonic final states via $L L e_c$ couplings.  Hence
the neutralinos must decay primarily to hadronic final states via couplings
like $L Q D$.  In this case, the cross section limits are about 0.6 pb
\cite{Expts}, and the sample point is not excluded.  Hence we expect decays
like $\chi^0 \rightarrow q\tilde{q}^* \rightarrow q q \ell$, so that the
final states for $\chi^0_2$ pair production will contain 4 jets together
with either charged leptons or missing energy if the $\chi^0_2$ decays
promptly.  If it decays first to $\chi^0_1 Z^*$ or $\chi^0_1
\tilde{\nu}_4^*$ there will be additional particles in the final state from
$Z,~{\tilde{\nu}} \rightarrow f \bar{f}$.  Conservative limits on the
dimensionless coefficients of $L Q D$ couplings among the first three
generations are at the level of 0.4-0.001 \cite{Dreiner}.  Under less
conservative assumptions, tighter constraints can be derived
\cite{Barbier}.  For instance, bounds from neutrinoless double beta decay
\cite{BabuHirsch} imply $\lambda'_{113} \lambda'_{131} < 8 \times 10^{-8}$.
Hence it seems not unlikely that the $b$-quark Yukawa coupling $L_4 Q_3
D_3$ is among the largest of the R-parity violating terms that can
contribute to neutralino decay.  If this is the case, the final states will
frequently include $b$-flavored hadrons.

\end{enumerate}

Although we have illustrated the qualitative features of this scenario for
a particular point in parameter space, certain predictions are fairly model
independent.  In particular, we expect that models of this type will always
possess at least one neutralino with mass below 65 GeV. Indeed, varying
$M_1$ and $M_2$ between 0 and 1 TeV, we find that the neutralino mass
matrix always has at least one eigenvalue below 65 GeV.  Also, we note that
the neutralino production cross section is not likely to be too small.  The
cross section from $Z$ exchange, summed over neutralino species, is larger
than 0.23 pb for all points in parameter space where $M_{1,2}\leq 1$ TeV,
where charginos cannot be pair produced at LEP-II, and where $Z$ decays to
pairs of neutralinos are kinematically forbidden.  The cross section could
be smaller if there are large cancellations between $Z$ exchange and
$t$-channel sfermion exchange. If we permit lighter charginos, then of
course we expect still more light particles and a richer phenomenology.  We
have emphasized the case of heavy charginos and lighter neutralinos since
constraints on R-parity violating charginos are quite stringent
\cite{Expts}.

{}We would like to end this note by summarizing some of the model
independent predictions of the above scenario.

\begin{enumerate}
\item
R-parity is necessarily broken by the Yukawa couplings.  As a result, the
production and decay of superpartners is modified compared to what one
finds in R-parity conserving models.  We have outlined the most novel
aspects of this scenario above in connection with neutralino pair production.
Apart from this, one expects the usual signatures of R-parity violating
supersymmetry, such as the absence of missing energy in SUSY particle
decays.

\item
A light Higgs is no longer a robust prediction of supersymmetry.
This is much in the spirit of \cite{KH} where it was pointed out that
$M_{Pf}$ suppressed operators can ameliorate an upper bound on the Higgs
mass in the non-supersymmetric context. In our case
the Higgs potential can come from the K\"ahler potential
\begin{equation}
 {X^*X \over M_{Pf}^2}(LL^* + {a\over M_{Pf}^2} (LL^*)^2+\dots)~,
\end{equation} 
which can in principle tolerate any mass the Higgs could have in 
the context of the non-supersymmetric Standard Model.

\item
Although we no longer necessarily expect a light Higgs, we do in general
expect one or more light neutralinos, with mass less than 65 GeV.  If the
charginos are too heavy to be pair produced at LEP-II, then at least one of
the neutralinos is likely to have a production cross section of 0.25 pb or
larger, which is not too far from present experimental limits.
\end{enumerate}

{}Finally, let us point out that the entire scheme discussed in this note
is possible because supersymmetry in the context of TeV scale gravity is
broken softly but (almost) completely. Thus, for instance, the Yukawa
couplings not allowed by supersymmetry are generated via higher dimensional
operators in the K\"ahler potential once supersymmetry is broken. On the
other hand, we have shown that the underlying supersymmetry leads to
non-trivial predictions in this scenario, such as the presence of light
neutralinos.  It is conceivable that this scenario can be confirmed or
excluded by LEP-II or the Tevatron.

{}We would like to thank Gia Dvali for collaboration at initial stages of
this work and valuable discussions. We would also like to thank Howard
Georgi, Henry Tye, and Cumrun Vafa for useful discussions.  The work of
A.K.G. was supported by NSF grant PHY-9802709.  The work of Z.K. was
supported in part by NSF grant PHY-96-02074, and the DOE 1994 OJI
award. This work was completed during Z.K.'s visit at the Enrico Fermi
Institute. Z.K. would also like to thank Albert and Ribena Yu for financial
support.

\end{document}